%% file: main.tex
\documentclass[journal]{IEEEtran}

\ifCLASSINFOpdf
  \usepackage[pdftex]{graphicx}
  \graphicspath{{./figures/}}
  \DeclareGraphicsExtensions{.pdf,.jpeg,.png}
\else
\fi
\usepackage[caption=false,font=footnotesize]{subfig}
\usepackage{amsmath}
\usepackage{algorithmic}
\ifCLASSOPTIONcompsoc
 \usepackage[caption=false,font=normalsize,labelfont=sf,textfont=sf]{subfig}
\else
 \usepackage[caption=false,font=footnotesize]{subfig}
\fi
\begin{document}
\bstctlcite{IEEEexample:BSTcontrol}
%
\title{Dot-Flik: A Scalable Edge AI Architecture for Distributed Insect Monitoring}%
%
%
%

\author{Mattia~Consani, Denisa-Andreea~Constantinescu, Åse~Håtveit, Titus~Venverloo, Fabio~Duarte, Carlo~Ratti, David~Atienza%
\thanks{Mattia Consani was with École Polytechnique Fédérale de Lausanne (EPFL), Lausanne, Switzerland (e-mail: mattia.consani@etik.com).}%
\thanks{Denisa-Andreea Constantinescu, and David Atienza are with École Polytechnique Fédérale de Lausanne (EPFL), Lausanne, Switzerland (e-mail: denisa.constantinescu@epfl.ch).}%
\thanks{Åse~Håtveit, Titus Venverloo, Fabio~Duarte and Carlo~Ratti are with the Massachusetts Institute of Technology (MIT), Cambridge, MA, USA (e-mail: aase@mit.edu; tvenver@mit.edu).}}


%
%

\markboth{Preprint --- Submitted for Publication}%
{Consani \MakeLowercase{\textit{et al.}}: Dot-Flik: Scalable Edge AI for Insect Monitoring}
%



\maketitle

\begin{abstract}
Global insect population declines necessitate scalable, continuous monitoring
systems, yet existing vision-based solutions remain constrained by high hardware
costs, energy demands, and reliance on centralized processing or cloud
connectivity. This article presents three contributions to address these
limitations. First, we propose a motion-informed frame filtering algorithm
based on temporal differencing, gamma-corrected motion amplification, and
block-based motion density analysis that discards irrelevant frames at the
edge while preserving insect activity, without requiring deep learning
inference on the sensing device. Second, we introduce a distributed,
hierarchical IoT architecture that decouples data acquisition from AI
classification through this edge-level preprocessing, projecting fractional
scaling of central processing requirements and significantly increasing
monitoring coverage compared to monolithic single-stream approaches.
Third, we validate the complete system through real-world outdoor deployments
on low-cost commodity hardware along four axes ---
real-time performance, network scalability, hardware cost, and energy
efficiency --- under varying wind conditions. Results demonstrate 60--80\%
frame reduction under light-wind conditions, sustained real-time
30~FPS operation with 12.8~ms of computational headroom, up to 22.6\%
energy savings, and support for 5--6 concurrent edge streams per central
node. These findings establish a practical foundation for dense, low-cost
biodiversity monitoring networks in urban environments.
\end{abstract}

\begin{IEEEkeywords}
Edge AI, insect monitoring, motion-based compression, distributed internet of things (IoT).
\end{IEEEkeywords}

%

\section{Introduction}
\input{chapters/01_introduction}

\section{Related Work}
\input{chapters/02_related_work}

\section{Dot-Flik System Architecture}
\input{chapters/03_system_architecture}

\section{Edge Motion-Based Data Reduction}
\label{sec:data_reduction}
\input{chapters/04_edge_motion-based_data_reduction}

\section{Experimental Setup}
\input{chapters/05_experimental_setup}

\section{Results}
\input{chapters/06_results}

\section{Discussion}
\input{chapters/07_discussions}

\section{Conclusion}
This work demonstrated that scalable, low-cost insect monitoring is achievable through principled decoupling of edge-level acquisition from centralised AI inference. Three complementary contributions realised this: (i) a motion-informed frame-filtering algorithm that eliminates irrelevant frames at the edge, achieving 60–80\% frame reduction at 30~FPS under light-wind conditions; (ii) a hierarchical Dot–Flik topology that reduces inter-node traffic sufficiently to support an estimated 5–6 concurrent sensing streams per classification node, establishing fractional scaling from a single-stream baseline; and (iii) outdoor field validation confirming up to 22.6\% energy savings and substantial computational margin for further development on commodity low-cost edge hardware.
More broadly, this work exemplifies a design principle for IoT vision systems: reducing data volume \emph{upstream} at the sensor, rather than optimising downstream processing, is the feasible and necessary path to network-scale deployment on commodity hardware. The edge-filter-then-infer pattern transfers naturally to any vision-based IoT workload where events of interest occur infrequently relative to total data capture---precision agriculture, wildlife observation, and infrastructure inspection---domains in which continuous full-resolution streaming is neither necessary nor feasible.
Future work will pursue adaptive threshold calibration to address sensitivity to dense foliage and high-wind deployment contexts, long-term seasonal field trials to assess operational stability beyond the evaluated conditions, and integration of an end-to-end classification pipeline within the validated architecture.


%



\section*{Acknowledgment}

The authors would like to thank the AMS Institute and the members of the MIT Senseable City Consortium. This work was partially funded by the UrbanTwin project, supported by the ETH Board's Joint Initiative program in the Strategic Area Energy, Climate and Sustainable Environment.

\ifCLASSOPTIONcaptionsoff
  \newpage
\fi

\bibliographystyle{IEEEtran}
\bibliography{bstctl,references}

%








\end{document}

%% file: chapters/01_introduction.tex
\begin{figure}[ht]
  \centering
  \vspace{-0.4cm}
\includegraphics[width=\columnwidth,trim=0.2cm 0 0.2cm 0,clip]{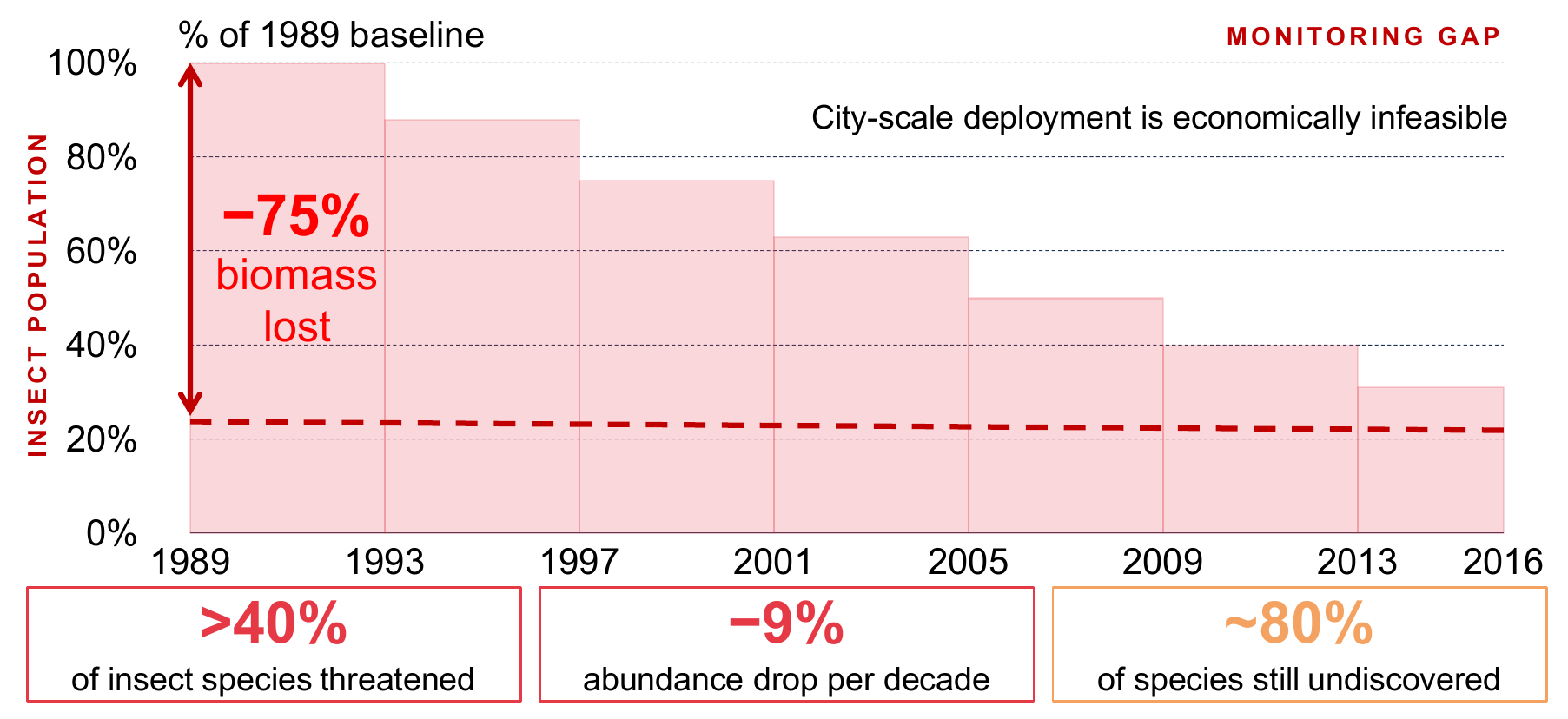}
    \vspace{-0.6cm}
  \caption{Motivation for cost-effective IoT insect monitoring solutions that scale. 
  }
  \label{fig:insect_decline}
\end{figure}

Insects represent the most diverse group of animals on Earth, comprising approximately half of all known multicellular species \cite{stork_how_2018, mora_how_2011, bar-on_biomass_2018}.
Alarmingly, global insect populations are currently facing a precipitous decline (Fig.~\ref{fig:insect_decline}), with some studies reporting biomass losses of up to 75\% in protected areas over the last few decades \cite{wagner_insect_2020, seibold_arthropod_2019, sanchez-bayo_worldwide_2019}. As insects play indispensable roles in terrestrial ecosystems, functioning as pollinators, decomposers, and foundational links in food webs, their preservation is critical for both ecosystem stability and agricultural productivity \cite{buxton_global_2022, thanuja_ecology_2025, jankielsohn_sustaining_2023}.
However, current understanding of these population dynamics is severely limited by traditional monitoring methodologies \cite{didham_interpreting_2020}.
Manual identification by entomologists is labor-intensive, slow, and suffers from low spatial resolution, making it unscalable for large-scale biodiversity assessment \cite{montgomery_standards_2021}.

Automated bio-monitoring represents a high-impact application domain for the Internet of Things (IoT), where dense networks of distributed sensors can provide continuous, \textit{in situ} observations of biodiversity at unprecedented spatial and temporal scales \cite{roy_towards_2024,van_klink_towards_2024}.
While recent advances in computer vision and deep learning have achieved high accuracy in species classification \cite{borowiec_deep_2022, beery_scaling_2021, tuia_perspectives_2022, weinstein_computer_2018, hoye_deep_2021, xia_insect_2018}, existing systems still face a fundamental barrier to scalability. Current state-of-the-art approaches typically depend on expensive, power-hungry hardware or continuous cloud connectivity \cite{bjerge_real-time_2022, preti_insect_2021}. These dependencies create major obstacles to large-scale deployment: high hardware costs constrain the number of nodes that can be installed, high power consumption often necessitates grid connectivity and limits placement flexibility, and cloud-centric architectures introduce network coverage constraints as well as privacy concerns in urban environments. As a result, scaling from a single monitoring site to a city-wide insect monitoring network remains both economically and technically prohibitive.

\begin{figure}[ht]
  \centering
  \vspace{-0.4cm}
\includegraphics[width=\columnwidth,trim=0.7cm 0 0.7cm 0,clip]{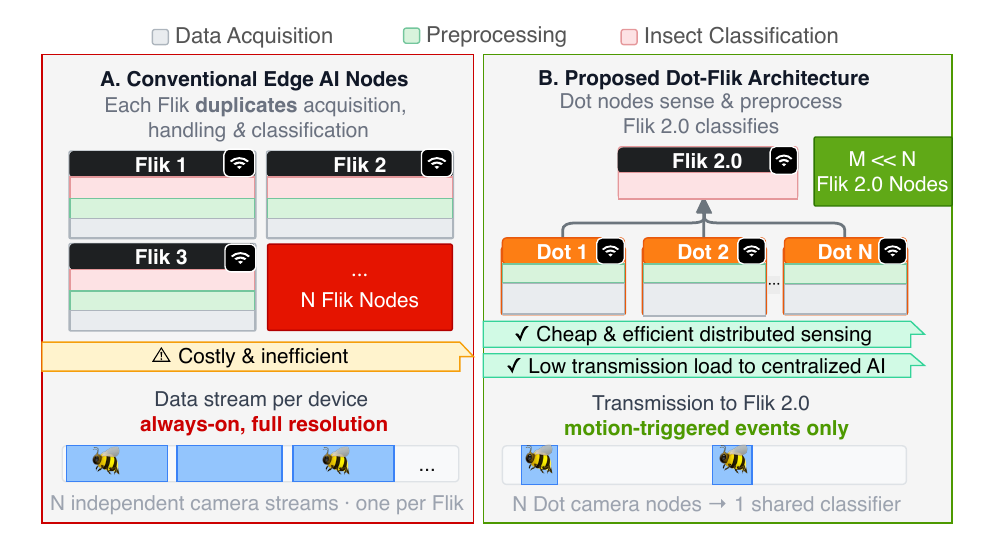}
    \vspace{-0.8cm}
  \caption{Conventional Edge AI camera nodes (A) duplicate acquisition, handling, and classification on every device. The proposed Dot-Flik architecture (B) splits sensing and preprocessing on Dot nodes from classification on Flik.}
  \label{fig:arch_intuition}
\end{figure}

Recent efforts have begun to address these challenges through edge-based architectures. In particular, the Sensing Garden project~\cite{senseable_lab} at Senseable City Lab developed \textit{Flik}, an edge AI platform capable of identifying numerous insect species under field conditions with high accuracy. The system decouples universal detection from species classification by implementing a staged on-device pipeline. Continuous data acquisition and handling process the full camera stream, while insect activity triggers the extraction of frame crops that are subsequently passed to the classifier, as illustrated on the left side of Fig.~\ref{fig:arch_intuition}.

Nevertheless, fundamental scalability limitations persist. Computationally, the current system operates near saturation, reaching 85\% CPU utilization with a single connected camera stream. Spatially, each device effectively covers only about 1~m$^2$, implying that broader area coverage ($N~m^2$) requires proportionally dense \textit{Flik} deployments. As a result, the compute, energy, and hardware cost per monitored area scale unfavorably, with each additional unit of coverage requiring additional sensing hardware, power provisioning, and backend processing capacity.

To overcome these limitations, this article proposes a distributed, hierarchical IoT architecture that decouples data acquisition from classification through edge-side data reduction, as illustrated on the right side of Fig.~\ref{fig:arch_intuition}. We introduce \textit{Dot}, a low-cost, energy-efficient sensing platform that performs lightweight, motion-informed frame selection at the source. Instead of transmitting continuous video and processing all captured frames at each \textit{Flik} node, the \textit{Dot}  filters background-dominated content locally and forwards only candidate insect observations to the edge AI node, \textit{Flik 2.0}. This separation of sensing and classification enables scalable and cost-effective distributed monitoring ($M << N$ expensive \textit{Flik 2.0} nodes and \textit{N} cost-efficient \textit{Dot} nodes) .

This article presents the design, implementation, and evaluation of a novel IoT architecture for distributed insect monitoring. The main contributions of this article are:
\begin{enumerate}
\item A \textit{lightweight motion-informed filtering algorithm} for sensing nodes that discards non-informative frames while preserving insect activity. This reduces the computational burden on the central processing units for edge AI and enables scalable distributed insect monitoring.

\item \textit{Dot-Flik}, a distributed, hierarchical IoT architecture that separates data acquisition from classification through sensor-level preprocessing. By assigning lightweight frame filtering to low-cost sensing nodes, the architecture reduces communication volume and enables backend edge-AI processing demand to scale fractionally with the monitored area covered by the sensing network, thereby improving the cost-effectiveness of large-area deployment.

\item \textit{Real-world outdoor deployment and validation} of the complete system, evaluated along four axes --- real-time performance, network scalability, hardware cost, and energy efficiency --- under varying weather conditions.
\end{enumerate}
The source code of the \textit{Dot} edge-node implementation is publicly available at \texttt{https://github.com/esl-epfl/dot-flik}.

The remainder of this article is organized as follows. Section~II reviews related work. Section~III describes the system architecture. Section~IV details the motion-based data reduction algorithm. Section~V presents the experimental setup. Section~VI reports the results, and Section~VII discusses the findings.

%% file: chapters/02_related_work.tex
Recent advancements in computer vision and deep learning have been pivotal in the development of automated insect monitoring systems. Hoye et al. \cite{hoye_deep_2021} argue that deep learning is destined to transform entomology by enabling non-invasive, continuous observation of insect abundance and behavior. Similarly, Preti et al. \cite{preti_insect_2021} reviewed camera-equipped traps, highlighting that while image sensors offer superior temporal resolution compared to human operators, they generate massive datasets that require automated processing. To address the challenge of species identification, Xia et al. \cite{xia_insect_2018} proposed an improved Convolutional Neural Network (CNN) strategy, utilizing a Region Proposal Network (RPN) and VGG16 backbone to enhance detection accuracy for multi-scale insect targets. Although these approaches demonstrate strong classification performance, they generally entail high computational costs that limit their applicability to resource-constrained IoT devices.

Building on classification capabilities, Bjerge et al. have advanced automated monitoring through integrated pipelines spanning automated light traps \cite{bjerge_automated_2021}, real-time insect-flower tracking \cite{bjerge_real-time_2022}, time-lapse deep-learning pipelines \cite{bjerge_deep_2024}, and most recently edge-side processing of camera-trap images \cite{bjerge_towards_2025}, with broader community efforts toward standardised AI-assisted monitoring frameworks \cite{roy_towards_2024}. However, these systems rely on post-hoc processing or substantial hardware resources to manage real-time data throughput, and none explicitly target the bandwidth and energy constraints of distributed IoT deployments.

To facilitate remote monitoring on constrained hardware, researchers have focused on dedicated IoT-oriented solutions. Suto \cite{suto_novel_2022} introduced a plug-in board designed for remote insect monitoring, emphasizing real-time data acquisition and low-power operation. Li et al. \cite{li_unified_2025} proposed a unified pest prevention system that leverages Social IoT concepts and a specialized lightweight model, GhostYOLOSA, significantly reducing computational load while maintaining high detection accuracy for small targets. Yang et al. \cite{yang_ehapzero_2025} introduced EHAPZero, a zero-shot learning framework that uses ensemble hierarchical attribute prompting to recognize unseen pest species on IoT devices without retraining. While these methods improve detection efficiency and flexibility on edge devices, they focus on the classification stage and do not address the upstream data redundancy problem: the energy wasted capturing and transmitting empty or irrelevant frames before inference even occurs.

The existing literature reveals a gap in balancing scalability with detection performance. The aforementioned methods---whether high-accuracy classifiers \cite{xia_insect_2018,bjerge_real-time_2022,bjerge_deep_2024}, lightweight models \cite{li_unified_2025}, or zero-shot frameworks \cite{yang_ehapzero_2025}---either stream full image data to the cloud or perform heavy on-device processing. Neither approach addresses the architectural optimization required for large-scale, distributed urban networks where bandwidth and power are strictly limited. This article addresses this gap by introducing \textit{Dot-Flik}, which relocates motion-informed filtering to a low-cost edge stage upstream of inference, so that backend processing demand scales with retained-frame volume rather than with the number of monitoring nodes.

%% file: chapters/03_system_architecture.tex
This section presents a hierarchical sensing and edge computing system designed to overcome the scalability limitations inherent in centralized insect monitoring solutions. The design comprises three distinct layers: (i) lightweight edge sensing nodes that perform motion-based data reduction, (ii) a central computing node that executes AI-driven classification, and (iii) an optional cloud backend for long-term archival and cross-site analysis when connectivity is available.
This configuration enables each edge sensor to contribute preprocessing capacity alongside raw data acquisition. The edge nodes are decoupled from the central processing node, allowing asynchronous operation at different processing rates. Specifically, edge nodes operate in real-time at the 30 FPS rate required by the \textit{Flik} classification pipeline, while the central device processes accumulated frames in batch mode to optimize computational efficiency.

\begin{figure}[ht]
    \centering
    \vspace{-0.4cm}
    \includegraphics[width=\columnwidth]{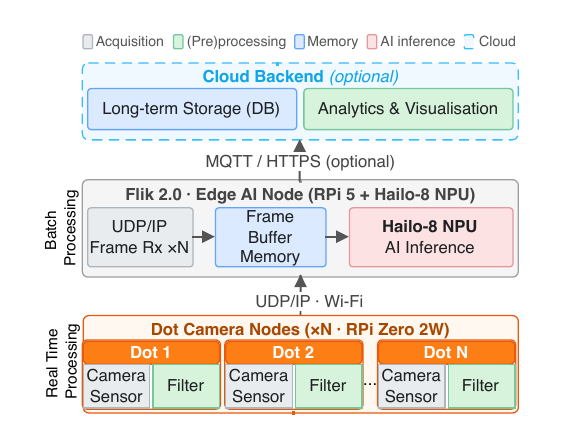}
    \vspace{-0.4cm}
    \caption{Dot-Flik system: N Dot nodes stream filtered frames via UDP/IP over Wi-Fi to a Flik unit; cloud backend optional.}
    \label{diag:sysArchitecture}
\end{figure}

The system design is governed by four key constraints that ensure practical deployability: (1) \textit{Real-time performance} for continuous video capture and processing at the edge without frame backlog; (2) \textit{Network scalability} to handle multiple concurrent data streams from distributed sensors without saturating the central processing node; (3) \textit{Low hardware cost} to enable dense spatial coverage across extensive monitoring areas; and (4) \textit{Energy efficiency} and autonomy for multi-day operation from commodity USB power banks (e.g., a 20{,}000~mAh unit), enabling deployment in locations lacking grid infrastructure.
Fig. \ref{diag:sysArchitecture} illustrates the Dot-Flik system for \textit{N} sensing nodes streaming filtered frames to a central classification unit. 

\subsection{Dot: Edge Sensing Node}
Edge nodes are implemented using Raspberry Pi Zero 2 W boards equipped with quad-core ARM Cortex-A53 processors and 512 MB RAM, paired with Raspberry Pi Camera Module v2 sensors. This configuration provides sufficient memory capacity for buffering multiple 640×640 pixel RGB frames while keeping per-node cost compatible with dense deployment. The platform's power efficiency enables battery-powered field deployment.
Each node implements a three-stage processing pipeline: (i) camera acquisition at 30~FPS---a rate dictated by the downstream insect tracking pipeline on the \textit{Flik} node, which maintains temporal identity of insects across consecutive frames---(ii) motion-informed frame filtering based on temporal differencing and block-based motion density analysis, and (iii) network transmission of filtered frames via hardware-accelerated H.264 streaming over Wi-Fi.
The motion-based data reduction technique is detailed in Section~\ref{sec:data_reduction}.

\subsection{Flik: Edge AI Insect Classification Node }
The central node comprises a Raspberry Pi 5 equipped with a Hailo-8 AI accelerator that executes detection and classification deep learning models on each received frame.
This node receives filtered streams from multiple edge sensors via dedicated network ports, storing frames to mass storage for asynchronous batch processing.
This temporal decoupling enables the system to process peak-period data during intervals of low insect activity, thereby maintaining near-constant computational utilization despite temporal variations in monitoring workload.
The central node additionally provides a local Wi-Fi network when external infrastructure is unavailable, ensuring communication in remote deployment scenarios. Alternative communication protocols may be substituted to accommodate region-specific requirements and infrastructure constraints.

%% file: chapters/04_edge_motion-based_data_reduction.tex
Motion-informed enhancement, originally proposed by Aguilar et al. \cite{aguilar_small_2022} for satellite video analysis and subsequently employed by Bjerge et al. \cite{bjerge_object_2023} for insect detection, represents a content-aware technique that this article adapts for edge-level frame filtering.
The lightweight design requires no deep learning inference on the edge device, enabling deployment on constrained hardware.

\begin{figure}[h]
  \centering
  \includegraphics[width=\columnwidth]{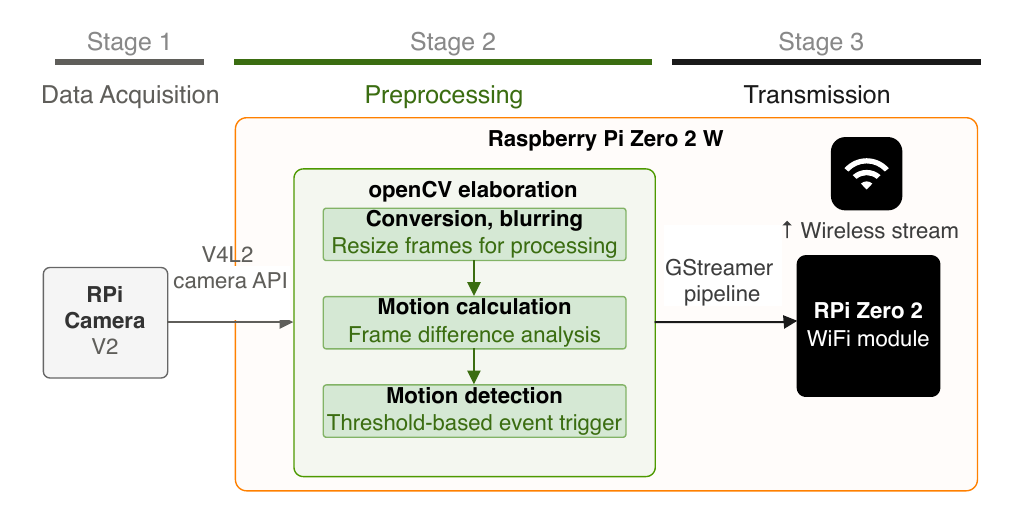}
  \vspace{-0.8cm}
  \caption{Dot workflow from camera acquisition to network transmission. 
}
  \label{fig:fullPipeline}
\end{figure}

The proposed approach leverages the principle that insects exhibit movement between consecutive frames, creating detectable temporal differences that distinguish them from static backgrounds. In outdoor deployments, this temporal signal is confounded by wind-induced vegetation movement, which generates dispersed, low-amplitude motion across the frame; the gamma correction and block-based density steps detailed below are designed specifically to suppress this class of background motion while preserving the localized signature of insect activity. The edge sensing platform implements a three-stage processing workflow: (i) data acquisition, (ii) motion-based frame filtering, and (iii) transmission of encoded frame streaming over Wi-Fi. This workflow is illustrated in Fig.~\ref{fig:fullPipeline} and formalized in Fig.~\ref{alg:frame_drop}. Stage (i) and (iii) involve standard camera capture and encoded frame streaming operations. The core contribution lies in stage (ii), which implements a motion-informed frame selection procedure that operates without deep learning models, enabling real-time execution on resource-constrained hardware.

The motion-based frame filtering stage (stage ii), formalized in Fig.~\ref{alg:frame_drop}, consists of three steps: (A) frame preprocessing, (B) temporal differencing and motion amplification, and (C) block-based motion detection. Each 640x640 RGB frame from stage (i) undergoes preprocessing operations to prepare it for motion analysis. First, grayscale conversion reduces the three-channel color image to single-channel intensity data. The frame is then downscaled to 320x320 pixels to reduce computational load. Finally, 5x5 Gaussian blur filtering suppresses high-frequency noise and attenuates micro-movements in vegetation that could trigger false motion detection.

For temporal differencing (step B), the system maintains a rolling buffer of three preprocessed frames to compute motion likelihood at each pixel location:
\begin{align}
    \Delta B_k[i,j] &= B_k[i,j] - B_{k-1}[i,j] \\
    L^{(3)}_k[i,j] &= |\Delta B_{k-1}[i,j]| + |\Delta B_k[i,j]|
\end{align}
where $B_k$ is the preprocessed frame at time $k$, $\Delta B_k$ the frame difference, and $L^{(3)}_k[i,j]$ the motion likelihood at pixel $(i,j)$ computed from the three most recent buffered frames; the superscript $(3)$ denotes the buffer size, not an exponent.

To enhance discrimination between concentrated insect motion and dispersed environmental noise (e.g.\ swaying vegetation under wind), the motion likelihood is normalized and gamma-corrected. The normalization divisor is $\max(\max(L^{(3)}_k),\,100)$, with the clamp at 100 preventing noise amplification in near-static scenes. Gamma correction with $\gamma = 3.0$ then emphasizes high-motion regions while suppressing background movement:
\begin{equation}
M_k[i,j] = 255 \cdot \left(\frac{L^{(3)}_k[i,j]}{\max(\max(L^{(3)}_k),\,100)}\right)^{\gamma}.
\end{equation}

In step (C), the motion-amplified frame is partitioned into non-overlapping 20x20 pixel blocks. The procedure computes the sum of motion intensities within each block using integral images, achieving $\mathcal{O}(1)$ computational complexity per block by eliminating nested loops for pixel summation.
For an integral image $I$, the sum of any rectangular region is calculated as:
\begin{equation}
    \begin{split}
        sum = I[y+h,x+w] - I[y,x+w] - \\
        I[y+h,x] + I[y,x]
    \end{split}
\end{equation}

A frame is retained for transmission if at least one block exceeds the motion threshold $\tau = 2000$. This block-based approach implements the concept of motion density—concentrated motion within specific blocks signals potential insect activity, while the same amount of motion dispersed across the entire frame indicates environmental noise such as swaying vegetation.
The high frame rate ensures insects crossing block boundaries will enter a block in subsequent frames, maintaining detection reliability despite spatial discretization. Frames failing to meet the threshold are discarded.

\begin{figure}[htbp]
\centering
\begin{algorithmic}[1]
\STATE \textbf{Constants:} $\gamma \leftarrow 3.0$,\ $K \leftarrow 20$,\ $\tau \leftarrow 2000$
\STATE \textbf{Input:} Video stream $\mathcal{S}$;\quad \textbf{Output:} Filtered stream $\mathcal{S}_{out}$
\STATE $\mathcal{B} \leftarrow \text{RollingBuffer}(3)$
\WHILE{System is Active}
    \STATE $F \leftarrow \text{Capture}(\mathcal{S})$
    \STATE $F_p \leftarrow \text{Blur}_{5\times5}\!\left(\text{Resize}_{320}\!\left(\text{Gray}(F)\right)\right)$;\ push $F_p$ to $\mathcal{B}$
    \STATE \textbf{if} $|\mathcal{B}| < 3$ \textbf{then continue}
    \STATE $L \leftarrow |\mathcal{B}_2 - \mathcal{B}_1| + |\mathcal{B}_3 - \mathcal{B}_2|$
    \STATE $T \leftarrow \max(\max(L),\,100)$
    \STATE $\mathcal{M} \leftarrow 255 \cdot (L / T)^{\gamma}$
    \IF{$\exists\; b \in \text{Blocks}_{K\times K}(\mathcal{M}):\; \textstyle\sum_{p \in b} p > \tau$}
        \STATE $\text{EncodeAndTransmit}(F)$
    \ELSE
        \STATE $\text{Drop}(F)$
    \ENDIF
\ENDWHILE
\end{algorithmic}
\caption{\textit{Dot} motion-informed frame-filtering algorithm.}
\label{alg:frame_drop}
\end{figure}

\subsection{Implementation}
Image processing uses OpenCV \cite{opencv} with OpenMP-parallelised routines and an integral image (\texttt{integral()}) for $\mathcal{O}(1)$ block-sum queries; camera capture and network streaming use GStreamer with Video4Linux (V4L2) and hardware-accelerated H.264 encoding.

Hyperparameters were set through prior-work adaptation and empirical tuning on field-deployment recordings. The gamma value $\gamma = 3.0$ and $5 \times 5$ blur kernel follow Bjerge et al.\ \cite{bjerge_object_2023}, with $\gamma$ raised from its original value via comparative testing for stronger contrast of concentrated insect motion. The normalization clamp of 100 was fixed empirically to suppress amplification of residual noise in near-static scenes, while the block size $K = 20$ and motion threshold $\tau = 2000$ were selected by systematic parameter sweep on field recordings to balance frame-rejection rate, retention of motion of interest, and the embedded compute budget.

\begin{figure}[htbp]
    \centering
    \includegraphics[width=\columnwidth, trim={0 25cm 0 10cm},clip]{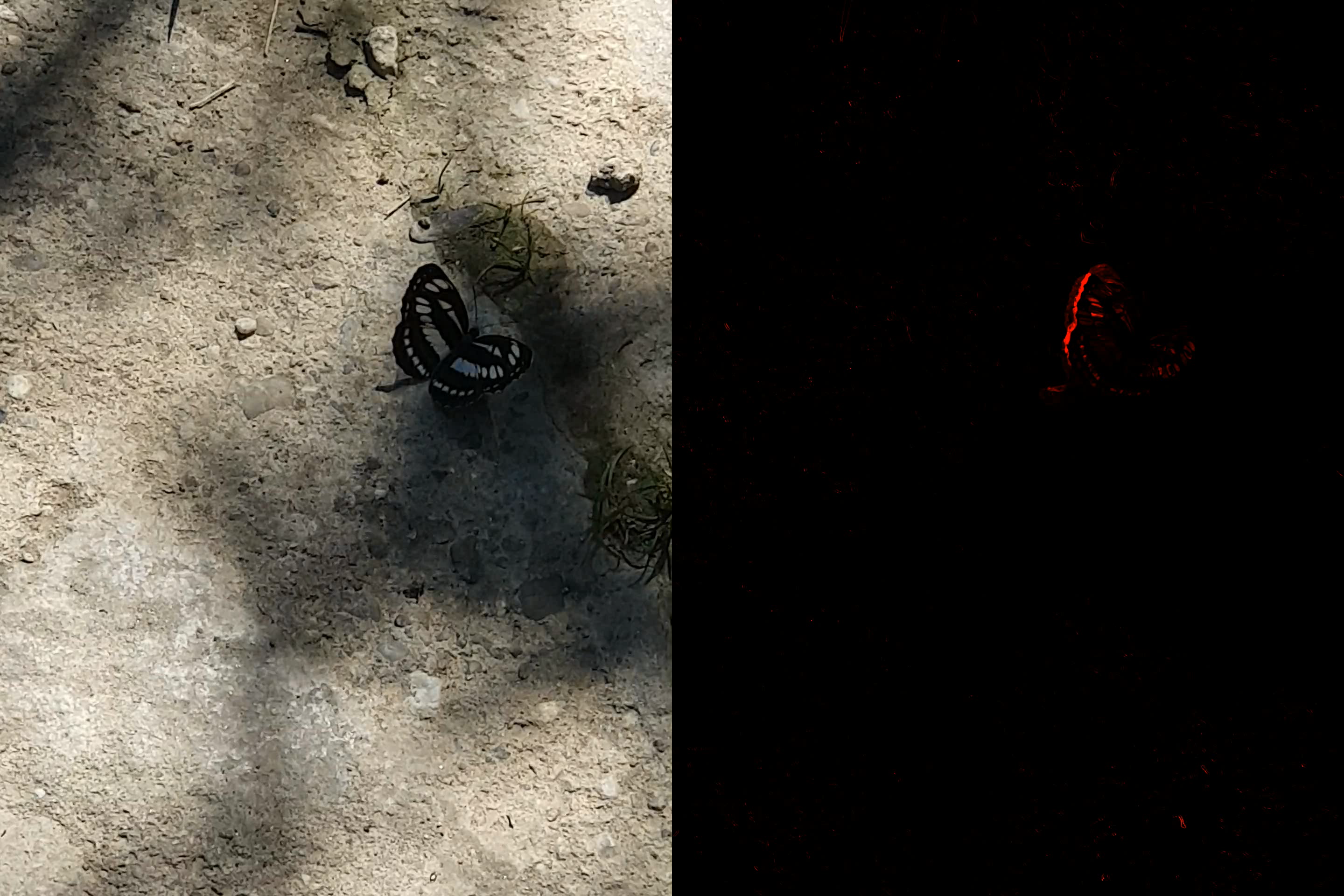}
    \caption{Motion-enhanced frame (right) and the last of the three consecutive original frames (left) used to compute it.}
    \label{fig:motion_enhanced}
\end{figure}

%% file: chapters/05_experimental_setup.tex
\begin{figure}[htbp]
    \centering
    \includegraphics[width=\columnwidth]{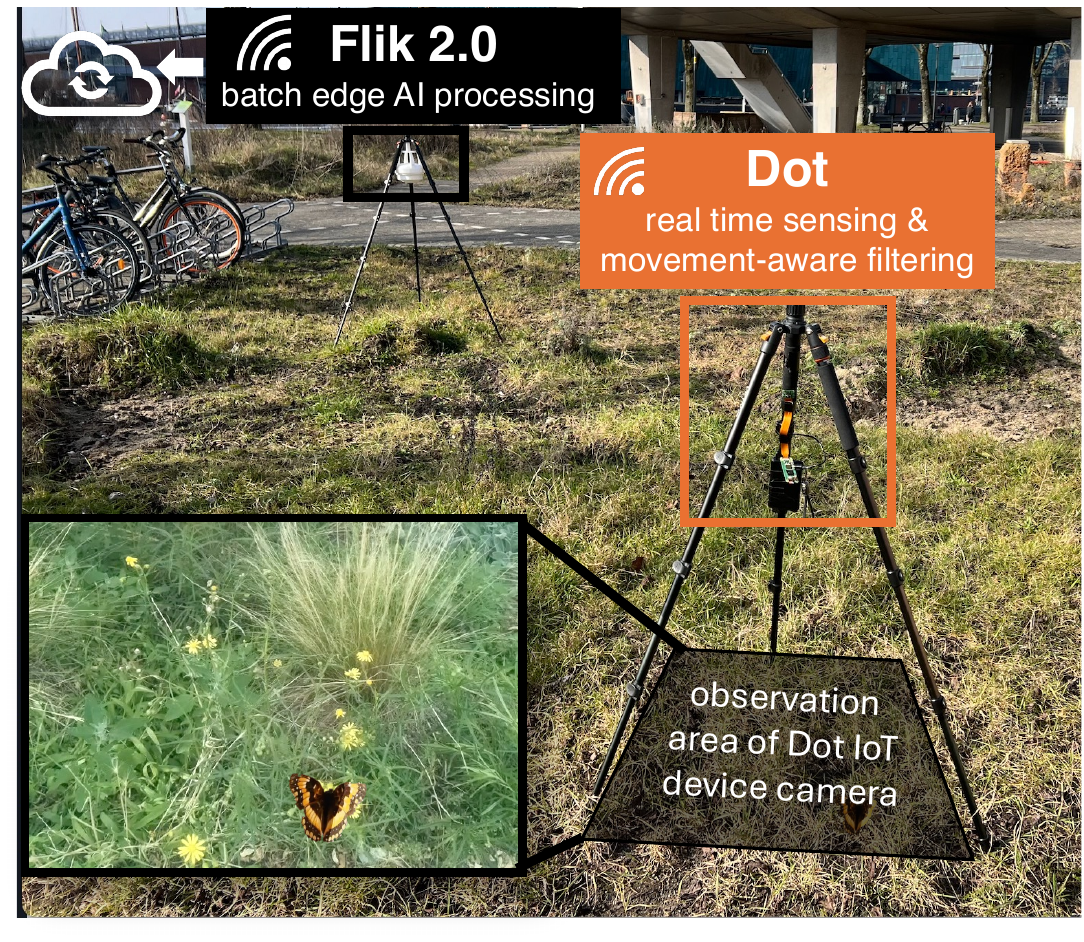}
    \vspace{-0.3cm}
    \caption{Experimental setup. \textit{Flik} and \textit{Dot} edge sensing platform mounted on tripods in the outdoor urban garden, showing camera field of view of the monitored surface.}
    \label{fig:deployment_setup}
\end{figure}

The system was evaluated in an outdoor urban garden environment with natural insect activity (Figure~\ref{fig:deployment_setup}). The presence of vegetation and flowers introduces realistic motion noise from wind-induced plant movement, providing challenging conditions representative of real-world biodiversity monitoring deployments.


\subsection{Evaluation Conditions}
Experiments were conducted under multiple environmental conditions to assess algorithm performance across varying wind intensities. Four wind categories were defined based on measured wind speeds: no wind (0--5~km/h), light wind (5--15~km/h), strong wind (15--35~km/h), and extreme wind ($>$35~km/h). Wind speed measurements were obtained from a local weather provider \cite{buienradar}. Each experimental run consisted of 20 minutes of continuous video capture and processing, generating sufficient data for statistical analysis while maintaining manageable dataset sizes.

Insect presence was not controlled; under high wind conditions insect activity naturally decreases \cite{karbassioon_responses_2023,Hennessy2020}, introducing variability in the number of insect-containing frames available for true positive rate calculation.

\subsection{Ground Truth and Evaluation Metrics}
Ground truth was established by manual verification of each transmitted frame, with insect activity confirmed visually as a qualitative check on the filtering approach. Performance was quantified across five metrics: frame drop rate, latency, CPU utilization, network throughput, and energy consumption. Latency was measured with embedded C++ \texttt{chrono} timers; CPU utilization with \texttt{perf}; network throughput with \texttt{iperf3} (maximum capacity) and \texttt{tcpdump} (application-level rates); and energy consumption with an inline USB power meter sampling at 1~Hz.

%% file: chapters/06_results.tex
\subsection{Data Reduction Performance}

\begin{figure}[htbp]
    \centering
    \includegraphics[width=\columnwidth]{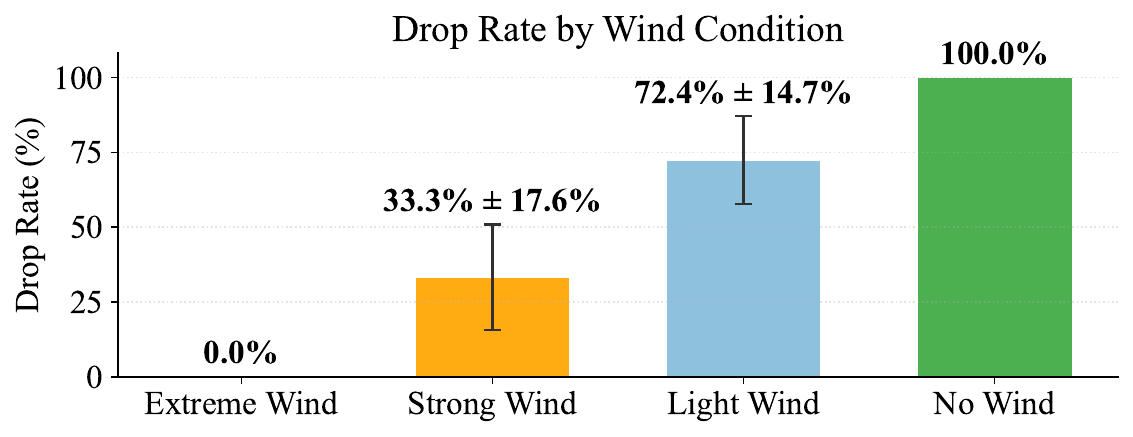}
    \caption{Frame drop rates across four environmental conditions, from extreme wind to calm; error bars reflect outdoor variability in wind, lighting, and vegetation motion.}
    \label{fig:data_reduction}
\end{figure}

The motion-informed frame selection algorithm achieved frame drop rates that varied systematically with environmental conditions, as illustrated in Fig.~\ref{fig:data_reduction}. Under light wind conditions (5--15~km/h), the system achieved 60--80\% frame reduction, while strong wind conditions (15--35~km/h) yielded 20--40\% reduction. Across all tested conditions, manual inspection of transmitted frames confirmed that insect activity visible in the unfiltered stream was consistently retained by the filtering algorithm. However, as insect presence was uncontrolled and the total number of observed events was limited, this observation is reported qualitatively rather than as a statistically validated detection rate.

Cross-referencing these drop rates with flying-insect ecology shows that the worst-case filtering regime --- strong wind (20--40\% frame reduction) --- coincides with suppressed insect activity \cite{karbassioon_responses_2023,Hennessy2020}, so it does not bound practical throughput. The binding scalability operating point is therefore light wind, $r = 0.724 \pm 0.147$, which together with the diurnal batch-processing model of the \textit{Flik 2.0} node governs the concurrent-stream projection developed in Section~VII.

\subsection{Real-Time Performance}

Table~\ref{tab:pipeline_performance} presents the timing analysis for each \textit{Dot} pipeline stage across three 20-minute profiling runs at maximum computational load (0\% frame drop rate).

\begin{table}[htbp]
\centering
\caption{Per-stage timing at 0\% drop rate (maximum computational load).}
\label{tab:pipeline_performance}
\resizebox{0.7\columnwidth}{!}{
\begin{tabular}{lcc}
\hline
\textbf{Pipeline Stage} & \textbf{Mean (ms)} & \textbf{Std Dev (ms)} \\
\hline
Camera Capture & 25.58 & 1.40 \\
Motion Processing & 6.22 & 1.03 \\
Streaming & 1.52 & 0.31 \\
\textbf{Total per Frame} & \textbf{33.32} & \textbf{1.07} \\
\hline
\end{tabular}
}
\end{table}

\begin{figure}[htbp]
    \centering
    \includegraphics[width=\columnwidth]{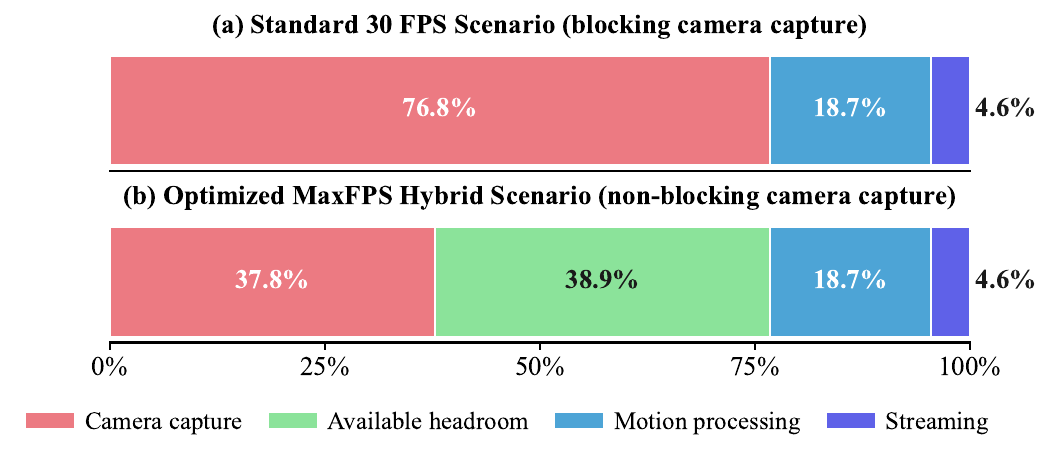}
    \caption{Per-frame time budget at 30~FPS (33.32~ms). (a) Standard 30~FPS. (b) Hybrid MaxFPS with non-blocking capture, showing 38.9\% headroom.}
    \label{fig:timing_breakdown}
\end{figure}

The system sustained 30 FPS operation with camera capture representing 76.8\% of total processing time, motion processing accounting for 18.7\%, and streaming 4.6\%, as shown in Fig.~\ref{fig:timing_breakdown}. Camera capture measurements include blocking mechanisms that enforce consistent frame rate delivery. Additional experiments conducted at maximum achievable frame rate (50 FPS nominal) revealed the unblocked camera capture time of 12.61 $\pm$ 2.08 ms, indicating 12.8 ms of available computational headroom within the 33 ms frame interval for further algorithm enhancements or additional processing tasks.

\subsection{Network and Storage Throughput}

Table~\ref{tab:io_performance} presents I/O performance measurements comparing maximum device capabilities with operational throughput at 0\% frame drop rate.

\begin{table}[htbp]
\centering
\caption{I/O Performance Analysis}
\label{tab:io_performance}
\resizebox{\columnwidth}{!}{
\begin{tabular}{lccc}
\hline
\textbf{I/O Component} & \textbf{Max Capability} & \textbf{Application} & \textbf{Utilization} \\
 & \textbf{(MB/s)} & \textbf{Throughput (MB/s)} & \textbf{(\%)} \\
\hline
Network Uplink & 2.81 $\pm$ 0.17 & 0.28 $\pm$ 0.04 & 10.0 \\
Network Downlink & 5.42 $\pm$ 0.55 & 0.28 $\pm$ 0.04 & 5.2 \\
USB Storage Write & 168.6 $\pm$ 1.52 & 0.15 $\pm$ 0.06 & 0.09 \\
\hline
\end{tabular}
}
\end{table}

The aggregate inflow at the \textit{Flik 2.0} node is bounded by its measured downlink capacity of 5.42~MB/s; per-Dot uplink at 2.81~MB/s is not binding, as each Dot's link is dedicated. With each Dot producing 0.28~MB/s of application traffic at 0\% drop rate, this implies a theoretical ceiling of $5.42 / 0.28 \approx 19$ concurrent unfiltered streams. This figure is an upper bound: IP/UDP/RTP framing overhead on the H.264 stream and multi-device WiFi medium arbitration both reduce the achievable aggregate throughput in practice, neither effect being captured by the single-device \texttt{iperf3} benchmark. Even after such discounting, the bandwidth ceiling remains well above the compute bound established in Section~VII-B. USB storage write bandwidth is negligible, confirming it is not a limiting factor for system scalability.

\subsection{Energy Consumption}

\begin{figure*}[h!]
    \centering
    \includegraphics[width=\linewidth]{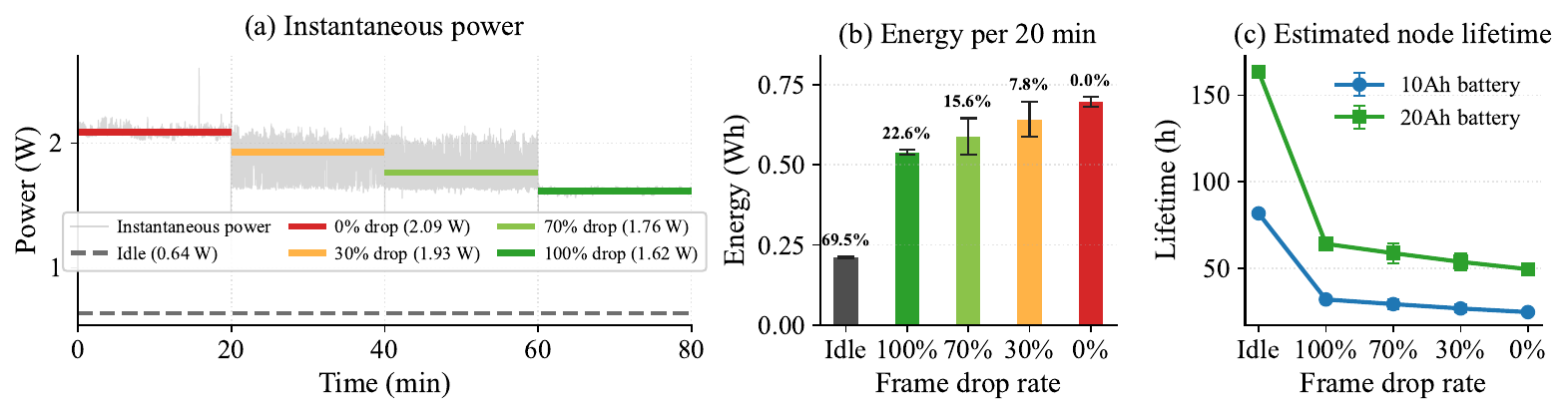}
    \caption{Power, energy, and lifetime versus frame drop rate. (a) Instantaneous power over four sequential 20-min intervals (0--100\% drop); horizontal bars mark interval averages, dashed line the idle baseline. (b) Energy per 20-min interval at each operating point, with savings annotated relative to 0\%. (c) Estimated node lifetime for 10 and 20\,Ah batteries.}
    \label{fig:energy_consumption}
\end{figure*}

Power consumption measurements across five operational phases demonstrated systematic correlation between frame transmission rate and energy draw (Fig.~\ref{fig:energy_consumption}). The system exhibited power consumption ranging from 1.6 to 2.1 W depending on frame drop rate. Energy savings increased progressively with drop rate: 7.8\% at 30\% drop rate, 15.6\% at 70\% drop rate, and 22.6\% at 100\% drop rate. Under moderate wind conditions (70\% drop rate), the measured energy reduction translates to 9.2 additional hours of continuous operation from a 20,000 mAh power bank, equivalent to extending deployment duration by approximately one day.

%% file: chapters/07_discussions.tex
\subsection{Interpretation of Results}

The block-based motion density approach reliably distinguished concentrated insect movement from dispersed environmental motion: despite substantial frame filtering, manual inspection confirmed that insect activity present in the unfiltered stream was consistently preserved, validating the filtering principle. 
The progressive energy savings with increasing drop rate (up to 22.6\%) confirm that motion-based filtering approach meaningfully extends field deployment viability beyond continuous streaming. A quantitative detection-rate measure would require controlled insect introduction or exhaustive annotation of unfiltered recordings, and remains future work.

The proposed content-aware filtering contrasts with the periodic downsampling (time-lapse) approach used in conventional insect camera traps. Bjerge et al.\ \cite{bjerge_towards_2025} compared time-lapse intervals from 10~s to 30~min against continuous tracking at 0.5~FPS and found that even the optimal 2-min interval introduced detection losses for rare taxa with fewer than 10 tracks per night. Because periodic sampling is content-blind, short-duration insect visits falling between capture intervals are irrecoverably lost. The motion-informed approach presented here avoids this trade-off by retaining frames based on detected activity rather than fixed timing, at the cost of reduced filtering effectiveness under high-wind conditions.

Wind sensitivity is the dominant deployment-context factor: dispersed vegetation motion under high wind generates false positives that reduce filtering effectiveness, while insect activity itself naturally decreases \cite{karbassioon_responses_2023,Hennessy2020}. Adaptive wind-aware thresholding could mitigate this.

\subsection{IoT Scalability Implications}

The distributed edge architecture addresses scalability limitations inherent in centralized processing models. Formally, each \textit{Dot} adds only $(1-r)F$ frames per second of central-node load during insect-active hours --- where $r$ is the edge-side drop rate and $F$ the acquisition rate --- against the full $F$ required per stream in the centralised baseline. Because flying-insect activity is diurnal, traffic arrives at \textit{Flik 2.0} only during daylight (approximately 12~h), while the node retains its 24~h processing budget; asynchronous batch processing therefore raises the effective active-hour ingest capacity to approximately $2F$, as the backlog accumulated in daytime is drained during idle nighttime intervals. The sustainable concurrency bound is consequently $n_{\max} = 2/(1-r)$. Substituting the measured light-wind filtering range ($r \in [0.60, 0.80]$ from Section~VI-A) yields $n_{\max} \in [5, 10]$ across the band, with the conservative end ($r = 0.60$) giving $n_{\max} = 5$ and the favourable end ($r = 0.80$) giving $n_{\max} = 10$. We adopt the conservative range of 5--6 concurrent \textit{Dot} streams per \textit{Flik 2.0} node as the design projection. This projection is derived from single-node measurements; validation with multiple concurrent physical nodes remains future work. Strong-wind regimes, where filtering degrades to 20--40\%, lie outside this operating envelope: flying-insect activity is simultaneously suppressed \cite{karbassioon_responses_2023,Hennessy2020}, and the resulting vegetation-driven traffic to \textit{Flik 2.0} carries negligible ecological signal. This regime therefore does not constitute a binding design bound; adaptive suppression of uninformative transmissions remains a direction for future work. The architectural advantage stems from distributed preprocessing at each edge node and asynchronous batch processing that smooths temporal variations in workload.


Network bandwidth is not the binding constraint. As established in Section~VI-C, the \textit{Flik 2.0} downlink admits approximately 19 concurrent unfiltered streams in the theoretical limit; even after discounting for protocol overhead (IP/UDP/RTP framing) and multi-device WiFi contention, this exceeds the 5--6 stream compute bound by a wide margin. Compute capacity at the \textit{Flik 2.0} node therefore governs scalability, not network or storage I/O, validating the decision to centralise inference rather than replicate it at each edge node.

Beyond capacity scaling, distributing \textit{Dot} sensors across plots also captures heterogeneous microhabitats, broadening the range of detectable species --- an ecological complement to the technical scalability benefit.

\subsection{Limitations}

This paper's scope is the \textit{Dot} edge-filtering layer and the \textit{Dot-Flik} architectural decomposition it enables; evaluation gaps concerning ground-truth methodology and monitoring-site diversity (absence of controlled insect introduction, single-deployment setting, limited environmental conditions) are inherited from the upstream \textit{Flik} classification pipeline, part of separate work by the Sensing Garden project, and fall outside this contribution. Within this scope, several constraints bound the generalisability of the reported results.
The experimental evaluation covered controlled wind variations; however, broader environmental phenomena---including precipitation, fog, and substantial illumination transients---were not systematically assessed. Seasonal variation in ambient light and vegetation density likewise falls outside the scope of this work, which focuses on the experimental validation of the Dot-Flik architecture and its scalability. Field deployment across diverse climates and extended time horizons remains an important direction for future evaluation.
The computational budget of the Raspberry Pi Zero 2 W constrains the algorithmic complexity of the Dot-node preprocessing pipeline. This is a deliberate design tradeoff: low per-unit cost and power consumption are prioritised over on-node intelligence, with heavier inference delegated to the centralised \textit{Flik 2.0} unit. More capable edge hardware would relax this constraint at the expense of deployment cost.
Finally, the fixed-threshold motion detection approach exhibits sensitivity to deployment context. In environments characterised by dense foliage or sustained high-wind exposure, spurious trigger rates increase, reducing filtering effectiveness.

\subsection{Future Work}
Adaptive motion-detection thresholding is the most pressing extension: integrating low-cost environmental sensors would allow the Dot node to dynamically recalibrate filtering parameters based on wind speed, illumination level, or time-of-day patterns, directly addressing the deployment-context sensitivity identified in this work.
Long-term seasonal field trials are needed to characterise classification accuracy drift, hardware reliability under thermal cycling, and the effect of seasonal vegetation changes on filtering effectiveness beyond the conditions evaluated here.
The architecture's decoupling of acquisition from inference makes it a natural substrate for federated learning: edge devices could share model gradients rather than raw frames, enabling collaborative retraining across distributed deployments while preserving data locality.
Finally, Wi-Fi was adopted as a convenient prototype-stage transport; investigating alternative connectivity options matched to the throughput, range, and power profile of specific deployment contexts remains future work.